\documentstyle[preprint,eqsecnum,aps,pra]{revtex}

\tightenlines
\begin{document}
\draft
\preprint{Nina-2}
\title{Non-linear effects in hopping conduction of single-crystal 
La$_{2}$CuO$_{4 + \delta}$ 
}
\author{B. I.  Belevtsev$^{*}$, N. V. Dalakova, A. S. Panfilov}
\address{B. Verkin Institute for Low Temperature 
Physics \& Engineering, \\ 
Kharkov, 310164, Ukraine
}

\maketitle
\begin{abstract}
\par
The unusual non-linear effects in hopping conduction of  single-crystal 
La$_{2}$CuO$_{4 + \delta}$ with excess oxygen has been observed.  The 
resistance is measured as a function of applied voltage $U$ 
(voltage controlled regime) in the temperature range 
5 K~$ \lesssim T \lesssim 300~$K and voltage range $10^{-3}$--25 V. 
At relatively high voltage (approximately at $U > 0.1$~V) the conduction 
of sample investigated corresponds well to variable-range hopping (VRH). 
Namely, in the range 0.1~V$ < U \lesssim 1$~V the conductivity does not depend 
on $U$ (Ohmic behavior) and the temperature dependence of resistance $R(T)$ 
follows closely the Mott's law of VRH [$R \propto \exp(T_{0}/T)^{1/4}$]. 
In the range of highest applied voltage the conduction has been non-Ohmic: 
the resistance decreases with increasing~$U$. This non-linear effect 
is quite expected in the frame of VRH mechanism, since the applied electric 
field increases the hopping probability. 
A completely different and unusual conduction behavior is found, however,
in low voltage range (approximately below 0.1~V), where the influence 
of electric field and (or) electron heating effect on VRH  
ought to be neglected.  
Here we have observed strong increase in resistance at increasing~$U$ at 
$T \lesssim 20$~K, whereas at $T > 20$~K the resistance decreases with 
increasing~$U$. The magnetoresistance of the sample below 20~K has been 
positive at low voltage and negative at high voltage.   
The observed unusual non-Ohmic behavior at low voltage range is 
attributable to inhomogeneity of the sample, and namely, to the enrichment 
of sample surface with oxygen during 
the course of the heat treatment of the sample in helium and air atmosphere 
before measurements. At low enough temperature (below $\approx 20$~K) 
the surface layer with increased oxygen concentration is presumed to 
consist of disconnected superconducting regions in poor-conducting 
(dielectric) matrix. This allows to explain the observed unusual 
non-linear effects in the conduction of sample studied. The results 
obtained demonstrate that in some cases the measured transport properties 
of cuprate oxides can not be attributed to the intrinsic bulk properties. 
\end{abstract}

\narrowtext
\section{Introduction}
\label{sec:Int}
\par
High-temperature (high-$T_{c}$) superconductivity of cuprate oxides with 
perovskite-related
structure is still fascinating problem in solid state physics. Aside from 
superconductivity the investigations of these materials give also the 
possibility of studying other fundamental phenomena, for example, magnetism, 
electron localization and hopping, metal-insulator transition. 
It is well known that electronic and magnetic properties of cuprate oxides 
depend essentially on charge carriers density which in its turn depends 
strongly on chemical composition. Introducing the donor or acceptor 
impurities into oxides, or changing the oxygen concentration in them it is 
possible to vary over wide limits their conductivity
and to cause in some cases the transition from insulating to metallic state. 
To judge whether a system is in the metallic, superconducting or insulating 
state the measurements of transport properties are used in most cases.
From these measurements the magnitude, temperature and magnetic-field 
dependences of resistivity and other conduction characteristics can be 
obtained. These data are often used for the characterization of prepared 
samples and evaluation of their ``quality''. But it is not uncommon that 
the measured transport characteristics do not correspond to the intrinsic 
crystal, stoichiometric and, therefore, electronic and magnetic properties 
of the sample. In the majority of the cases the main reason for it is  
a sample inhomogeneity 
due to peculiarities of sample preparation procedure, heat treatment and 
other related factors. Two main possible sources for cuprate 
inhomogeneity can be distinguished: 
intrinsic and extrinsic. Intrinsic source is connected with phase separation 
of cuprate oxides on two phase with different concentration of charge 
carriers \cite{ps}.  The extrinsic one is due to various technological 
factors of sample preparation. This may lead, among other things, to 
significant difference in  charge carriers density between the surface and 
inner parts of the sample \cite{browning}. 
\par
In our opinion the investigations of influence of surface or volume 
inhomogeneity of cuprate oxides on their transport properties are of 
considerable importance in two following aspects. Firstly, such kind of 
studies can help to answer the question in what degree the observed 
transport properties can be attributed to intrinsic properties of bulk 
crystal \cite{browning}. Secondly, under gaining enough experimental 
data on this matter (combined with necessary theoretical considerations 
and treatments) it is possible to apply the transport measurements 
not only for revealing of structural inhomogeneity, but also for 
identification of specific types of surface and volume inhomogeneities. 
Therefore, study of influence of inhomogeneity on transport properties of 
cuprate oxides is of both fundamental and applied importance.
\par
In this communication we shall describe some new results of 
investigation of hopping conduction of single-crystal 
La$_{2}$CuO$_{4 + \delta}$ with some amount of excess oxygen 
($\delta \not= 0$). In the studies of high-$T_{c}$ superconductivity 
these compounds have attracted considerable attention. The 
stoichiometric  La$_{2}$CuO$_{4}$ ($\delta = 0$) is an antiferromagnetic 
insulator with N\'eel temperature $T_{N}$ in the range 300-325~K 
\cite{ps,john,birg}. However, the introducing of excess oxygen (as well as 
doping with bivalent metals such as Sr) leads to the violation of  
long-range antiferromagnetic order and to a transition to  metallic 
and superconducting states. Excess oxygen doping introduces additional 
charge carriers (holes) in Cu-O planes \cite{kremer}. For high doping level 
($\delta \simeq 0.13$) the superconducting transition temperature 
$T_{c}$ may be as high as $\simeq 50$~K \cite{sato}.  
For the range of doping $\delta \simeq 0.01- 0.055$ the 
La$_{2}$CuO$_{4 + \delta}$ compounds undergo a phase separation below room 
temperature into the two phases with different oxygen content: the 
oxygen-poor phase is nearly stoichimetric and non-superconducting, while 
oxygen-rich phase is superconducting \cite{ps,kremer,jorg,hund,ryder,zakh}.  
Depending on $\delta$ value the different (sometimes coexisting) 
superconducting phases can emerge with $T_{c}$ values from 
$\simeq 20$~K to $\simeq 45$~K \cite{ps,kremer,jorg,hund,ryder,zakh,katya}. 
\par
It is known that low-temperature conduction of nearly stiochiometric 
La$_{2}$CuO$_{4 + \delta}$ occurs by variable-range hopping (VRH) of 
localized holes \cite{kast,zakh2,bel1,bel2} and can be fitted well to 
Mott's formula [with temperature dependence of resistance 
$R \propto \exp (T_{0}/T)^{1/4}$]. 
In Ref. \cite{bel1,bel2} it was found that the transition from VRH to 
simple activation conduction [$R \propto \exp (\Delta/kT)$] occurs at 
temperatures below 20 K (the similar effect was described also before in 
Ref. \cite{zakh2}). In Ref. \cite{bel1,bel2} this effect was explained 
by the influence of sample inhomogeneity, namely, by presence of 
superconducting inclusions in insulating sample due to phase 
separation of La$_{2}$CuO$_{4 + \delta}$.  
As was mentioned in Ref. \cite{browning}, in each case when some 
exotic transport behavior of cuprate oxides is found, the reason for it 
should be sought primarily in the possible influence of inhomogeneity. Our
new experimental results support (as we believe)  this point of view. 
We have observed the unusual non-linear behavior of hopping conduction 
of single-crystal La$_{2}$CuO$_{4 + \delta}$: at low applied voltages 
(in conditions where the influence of electric field and (or) electron 
heating effect on VRH can be neglected) the resistance strongly increases
with increasing of applied voltage at $T \lesssim 20$~K, but decreases with 
voltage increasing at temperatures above 20 K. This unusual non-Ohmic 
behavior is attributed to inhomogeneity of sample, and namely, 
to the enrichment of sample surface with oxygen during 
the course of the heat treatment of the sample in helium and air atmosphere 
before measurements. At low enough temperature (below $\approx 20$~K) 
the surface layer with increased oxygen concentration is presumed to 
consist of disconnected superconducting regions in poor-conducting 
(dielectric) matrix. 
 
\section{Sample and experiment}
\label{sec:Sample}
\par
We have studied the hopping conduction of same single-crystal sample of 
La$_{2}$CuO$_{4 + \delta}$ as in Ref. \cite{bel1,bel2}, but with reduced
and inhomogeneous oxygen content in it as a result of outlined below heat 
treatment in 
helium gas and air. The original sample or it is better to say the original 
state of this sample is characterized by $T_{N} \simeq 230$~K and 
$\delta \approx 0.005$ \cite{bel1,bel2}. For reducing the oxygen content 
the sample was annealed in a furnace in an atmosphere of 
helium at $T \simeq 330^{\circ}$C for two hours. It was cooled thereafter 
rather slowly (about 4 hours) in the same inert atmosphere. It is known that 
annealing in inert gas atmosphere is very effective way to reduce the 
oxygen content in cuprate oxides \cite{kast}. The resistance of the sample 
after this procedure (measured by a standard four-probe technique) 
has appeared however to be too high (about 1.7 k$\Omega$) for intended 
study of hopping conduction at low temperatures (down to about 5~K). 
Therefore it was additionally annealed in air at $T \simeq 330^{\circ}$C 
for 2.5 hours and (for lowering the contact resistance)  
at $T \simeq 80^{\circ}$C for two hours. In the result of such heat 
and gas treatment the oxygen content in single-crystal sample was 
reduced significantly. This is evidenced by increase of N\'eel temperature 
$T_{N}$ from 230 K to 290 K and very large rise (on more than three order of
magnitude at liquid helium temperatures) in resistance (see Figs. 1 
and 2 in which the temperature dependences of magnetic susceptibility  
and resistance are shown for the original state of the sample and for the 
state after above-described gas and heat treatment). The rather high
crystal quality of sample investigated is characterized by high anisotropy
of magnetic susceptibility (Fig.~\ref{Fig.1}).   
\par
The sample studied has dimensions approximately $3\times 3 \times 2$~mm.
For resistance measurement the thin gold contact wires were connected to
the sample by a silver epoxy paste which was hardened at 
$T \simeq 80^{\circ}$C for two hours. The measuring direct current $I$ 
was parallel to the Cu-O planes. Two techniques were used in resistance 
measurements: (i) A standard four-probe technique when sample
resistance was less than $\approx 4\cdot 10^{6}$~$\Omega$; (ii) Two-probe 
technique for higher sample resistances. For both techniques, actually,  
the $I-V$ characteristics were measured with  applied voltage varying 
$U$ (voltage controlled regime). From these data we will present below 
the obtained dependences $R(T,U)$ and $I(U)$. 
\par
In measurements of high-resistive semiconducting samples with non-Ohmic
effects it is important to take into account the possible influence of 
contact resistances.  Concerning our sample, we can say 
the following on this point. Firstly, both of the (four-probe and two-probe) 
techniques give the same behavior of $I$-$V$ curves and quite closely 
values of resistance $R = U/I$ (as a rule, the difference is not more than 
about 2 \%) in the resistance range $2\cdot 10^{5}$--$4\cdot 10^{6}$~$\Omega$. 
This range corresponds approximately to temperature 
range 15--25 K. Secondly, the special estimation of influence of contact 
resistances (using different contact places or short-circuiting wires) at 
$R \geq 10^{9}$~$\Omega$ (this corresponds to temperature below 10~K)
has shown that the ratio of contact resistance to measured sample 
resistance is less than 10~\%.  All this implies (as we believe) that the 
contact resistance has not much influence on reliability of results 
obtained.
\par 
The $I$-$V$ curves and resistance  were also recorded in magnetic field 
$H$ (with magnitude up to 5 T) in the temperature range 5-40 K. The 
magnetic field was directed along the Cu-O planes at right angle to the
measuring current. 

\section{Results and discussion} 
\par
We found  that hopping conduction of sample investigated 
follows closely the Mott's law of VRH:
\begin{equation}
R \propto \exp \left( \frac{T_{0}}{T} \right)^{1/4},
\label{one}
\end{equation}
where $T_{0} \simeq 6.4\cdot 10^6$~K. It can be seen in Figs. 2 and 3 that 
this law holds for broad temperature range (10-300 K) in which the 
resistance is varied up over 7 orders of magnitude. The same exponential 
$R(T)$ dependence in nearly stoichiometric La$_{2}$CuO$_{4 + \delta}$ was 
found previously  in Refs. \cite{kast,zakh2,bel1,bel2}, but in not as wide 
temperature and resistance ranges as in present study. In the theory of 
VRH the fractional exponent in Eq.\ (\ref{one}) is written in general form 
as $\alpha = 1/(D+1)$ where $D$ is system dimensionality \cite{mott}. 
Therefore, the case $\alpha = 1/4$, observed in our and previous studies,  
corresponds to behavior of a three-dimensional system. This seems to be 
contrary to the commonly accepted
belief that the  cuprate oxides with layered perovskite structure in which 
the Cu-O planes are main conducting units should behave as electronic quasi 
two-dimensional systems  \cite{iye,forro,loktev}. If this is the case, 
the VRH behavior should be two-dimensional with $\alpha = 1/3$, and this was 
indeed observed in some cuprate oxides \cite{iye,forro}. However, as shown in 
Ref.\cite{kremer}, in  La$_{2}$CuO$_{4 + \delta}$ owing to special 
character of the excess oxygen as interstitial atom with weak oxygen-oxygen 
bonding a hole transfer between Cu-O planes is likely. Therefore, the VRH of
this compound behaves as that of three-dimensional system. In passing it 
should be mentioned that this is not true for Sr or Ba doped 
La$_{2}$CuO$_{4}$ systems which remain quasi two-dimensional \cite{kremer}.  
\par
At $T \lesssim 20$~K, we observed very large deviations of $R(T)$ 
dependence from Mott's law (Fig.~3) which are determined by 
non-Ohmic effects in the sample conduction. In this temperature 
range the resistance rise with decreasing temperature is much less than 
the prediction of Eq.\ (\ref{one}) and at low enough 
temperatures the resistance does not increase at all [approaches some 
constant value or even decreases with decreasing temperature at fairly 
low voltage (Fig.~4)]. The deviation temperature  below which the appreciable 
deviations of this type take place 
decreases as the voltage $U$ increases. With this a quite unusual and 
unexpected behavior for semiconductor in VRH regime of conduction  is 
connected: at low enough temperature ($T \lesssim 20$~K) the resistance 
{\it increases} with $U$ increasing (Fig.~3). Indeed, it is well known 
\cite{mott} that 
conductivity in this regime can only increase with the applied electric
field $E$ which (when it is large enough) enhances the electron hopping
probability. For not very large $E$ values ($eEL_{c} < kT$, 
where $L_{c}$ is the localization length), the effect of electric field 
on resistance can be described by the following expression \cite{mott}:
\begin{equation}
R(T,E) = R_{0}(T) \exp \left(- \frac{eEr_{h}\gamma}{kT} \right),
\label{two}
\end{equation}
where $R_{0}(T)$ is the resistance for $E \rightarrow 0$ 
[described by Eq.\ (\ref{one})], $r_{h}$ is the mean hopping distance, and
$\gamma$ is a factor of the order of unity. It follows from Eq.\ (\ref{two})
that at low enough field ($E \ll kT/eL_{c}$) the resistance does not
depend on $E$ that is the Ohm's law holds. With increasing $E$ and 
decreasing $T$ the influence of electric field must be enhanced and lead 
to decreasing in $R$ with increasing $E$ that is quite contrary to that 
we have observed (Fig.~3).
\par
The described unusual $R(U)$ behavior is one of the major 
non-Ohmic effect which we have observed. Before trying to explain it 
we should, however, present more general picture of found non-Ohmic effects 
in $I$-$V$ characteristics and corresponding $R(U)$ dependences 
of sample investigated (Figs.~5--7).  At low enough voltage the 
resistance behavior was found to be non-Ohmic in all temperature range 
investigated (from 5 K to room temperature), but at $T \lesssim 20$~K the 
resistance increases with increasing $U$ (as was shown above)  
whereas at $T \gtrsim 20$~K it decreases with increasing $U$ 
(Figs.~6 and 7).{\Large{\footnote {Owing to logarithmic scales 
in Fig. 6 it can not be seen the important peculiarities of the 
$R(U)$ behavior at temperatures above 100 K. Because of this, some examples 
of $R(U)$ dependences in this temperature range are shown more clearly in 
Fig. 7 using semilogarithmic coordinates.}}} 
These unusual $R(U)$ dependences at low voltage and radical difference 
between them below and above $T \approx 20$~K are keys to understanding 
of conducting state of sample investigated and will be considered more 
thoroughly below.  At higher voltage the  $I(U)$ and $R(U)$ behaviors are  
basically the same for all temperature range investigated. Namely, in 
some intermediate range of voltage the Ohm law is true and at maximal 
upplied voltage (about 10 V or more) the resistance decreases with 
increasing $U$ (Figs. 5--7). As it was mentioned above, this type of 
transition 
from Ohmic to non-Ohmic regime of conduction at increasing of applied 
voltage is quite common for semiconductors with VRH and is attributed to 
the influence of applied electric field \cite{mott}.  We believe that this 
is also true for the sample studied and can substantiate it with some 
numerical estimates using the Eq.\ (\ref{two}). Indeed, it is known 
\cite{kast,chen} that electron localization length $L_{c}$ in nearly 
stoichiometric La$_{2}$CuO$_{4 + \delta}$ is about 0.8--1.0 nm. Taking into 
account that the mean hopping distance $r_{h}$ in VRH regime of conduction 
is greater than $L_{c}$ (say by a factor of 2 or 3), and using the 
above-indicated sample dimensions, it is easy to see that the electric field 
effect on hopping conduction is  negligible ($eEr_{h}\gamma/kT \ll 1$)
not only in low-voltage range where the above-mentioned 
non-linear $I(U)$ behavior and unusual $R(U)$ dependences were observed,
but also in higher-voltage range, where Ohmic behavior takes place.
Only for the highest applied voltage (10 V and more) the quantity 
$eEr_{h}\gamma/kT$ may be about 0.1 and, hence, the influence of electric 
field $E$ in accordance with Eq.\ (\ref{two}) can be appreciable. 
This can explain the resistance decrease with U increasing at highest 
applied voltage (Figs. 5--7). Besides, at fairly high field the heating 
effect is possible at low temperatures. 
This can also lead to the resistance decrease with $U$ increasing.    
\par
The magnetoresistance (MR) of sample studied was found to have 
appreciable and rather high magnitude only below $T \simeq 10$~K. It was
negative at high voltage range, but at low voltage ($U \lesssim 0.1$~V)
the MR becames positive at low enough temperatures (Figs. 8 and~9).
The negative MR is quite common for insulating La$_{2}$CuO$_{4 + \delta}$ 
samples  and may be determined by different mechanisms 
\cite{zakh2,bel1,bel2,thio}, which we will not discuss here in detail. 
As far as we know, the positive MR in insulating La$_{2}$CuO$_{4 + \delta}$ 
was not observed. Theoretically this phenomenon is considered, however, 
as quite 
possible in the VRH regime of conduction and is associated with the 
shrinking of the impurity wave function in a magnetic field \cite{shklo}. 
In this connection we have calculated the 
possible value of MR using the appropriate equation in Ref. \cite{shklo}
for the case of ``weak'' magnetic field ($L_{H} \gg L_{c}$, where 
$L_{H} = (\hbar/eH)^{1/2}$ is the magnetic length):
\begin{equation}
\ln \frac{R(H)}{R(0)} = 
t_{1}\left (\frac{L_{c}}{L_{H}}\right )^{4} 
     \left (\frac{T_{0}}{T} \right )^{3/4},
\label{three}
\end{equation}
where $t_{1} = 5/2016$. We have obtained that
$\ln [R(H)/R(0)]$ is about 0.003 for $H = 4$~T. This is much less 
than the experimental value of $\ln [R(H)/R(0)] \simeq 0.2$ (Fig. 9). We
believe, therefore, that the observed positive MR is not determined by the
mechanism proposed in Ref. \cite{shklo}.
\par
From the above discussion it appears that the conduction behavior of 
sample studied at high voltage range, in particular, the transition from
Ohmic to non-Ohmic regime of conduction with $U$ increasing, is quite 
consistent with known properties of semiconductors. This is not the
case, however, for the observed non-linear effects in low-voltage
range. This raises the two main questions: (i) Why this non-Ohmic effects
take place at all at so low voltage [in conditions where the influence of 
electric field and (or) Joule heating on VRH can be neglected]? 
(ii) What is the cause of radical
difference between non-Ohmic effects below and above $T \approx 20$~K
in this voltage range? What is more, the observed transition from negative 
to positive MR at decreasing $U$ should also be considered. 
After examination of obtained results   and taking into account the known 
properties of cuprate oxides 
we arrive at the conclusion that the sample inhomogeneity, namely, surface 
enrichment with oxygen may be responsible for the observed 
non-linear conduction effects. For the rest of the paper we will present 
the points substantiating this conclusion. 
\par
First of all we would like to point out that as the result of the 
above-described heat treatment of the sample in helium and air 
(see Sec.\ \ref{sec:Sample}) the oxygen concentration at the surface of 
sample may be considerably higher than in central (inner) region of it. 
Indeed, the first step of the treatment was an annealing in helium gas. 
This should cause \cite{kast} the effective reducing in oxygen content 
in the sample. However, the second step was an annealing in air (partly 
for the purpose of reducing in contact resistance) and this could definitely 
cause the oxygen enrichment of sample's surface region. This is quite 
possible if after the helium treatment the oxygen concentration in the 
sample was low enough. Consider in this connection once again the temperature 
dependence of magnetic susceptibility $\chi(T)$ of sample studied (Fig.~1). 
It can be seen that after the described heat and gas 
treatment the N\'eel temperature has increased from $\approx 230$~K to 
$\approx 290$~K. The later value of $T_{N}$ corresponds to nearly  
stoichiometric La$_{2}$CuO$_{4 + \delta}$ (very low oxygen content). 
Therefore, the heat treatment in helium was fairly effective in reducing the 
of oxygen concentration. At the same time, if a considerable volume part 
of the sample has gained some additional oxygen after the heat treatment
in air, this should be reflected in the form of the $\chi(T)$ curves as 
well. However, any marked evidence of sample inhomogeneity in this curves,
can not be seen. There is only one distinct 
peak in $\chi(T)$ dependence. But it should be taken into account that in 
the case, when only fairly thin surface layer has increased oxygen 
concentration, the influence of it on $\chi(T)$ dependence may be quite 
negligible. It should be also noted that the marked difference between the 
surface and inner oxygen content is rather common for the 
La$_{2}$CuO$_{4 + \delta}$ and other cuprate oxides \cite{browning,sato}.  
For example, in Ref.~\cite{sato} in La$_{2}$CuO$_{4 + \delta}$ films, which 
were oxidized in ozone gas, the increased oxygen concentration in surface 
layer was found. Taking all this into account and considering the 
peculiarities of sample treatment we shall conceive in the following
that the surface region of the sample is enriched with oxygen.{\Large{\footnote 
{Below will be considered alone the possible influence of this type of 
inhomogeneity on the conduction. We exclude the phase 
separation in inner part of sample as other source of inhomogeneity. Indeed,
the N\'eel temperature $T_{N} \approx 290$~K for the sample studied 
means that $\delta \leq 0.003$ \cite{zakh,birg2}. This value of $\delta$ 
is far outside of $\delta$ range (between $\delta \simeq 0.01$ and 
$\delta \simeq 0.055$) in which the phase separation occurs 
\cite{ps,kremer,jorg,hund,ryder,zakh,birg2}.}}}  Based on this, it is 
possible to give the reasonable explanation of all obtained results.
\par
The oxygen-enriched surface layer of sample can undergo a phase separation 
\cite{ps,kremer,jorg,hund,ryder,zakh} with the resulting formation of 
considerable volume fraction of superconducting phase. In this case the
surface layer would consist of disconnected superconducting regions in
poor-conducting (dielectric) matrix. We believe, that critical temperature 
$T_{c}$ of superconducting phase is about 20 K in the case being considered. 
It is at this temperature that the radically change in non-linear behavior 
of conduction takes place (Figs.~5 and 6).{\Large{\footnote 
{The stable superconducting phase with $T_{c} \simeq 20$~K can emerge due to
phase separation of La$_{2}$CuO$_{4 + \delta}$ at rather low oxygen doping 
level ($\delta \simeq 0.01$) \cite{katya}.}}}  Consider, at first, the 
conduction below $T \simeq 20$~K. In the specified conditions, for driving 
electric field the system provides at least two main channels for the 
response: the low-resistive surface layer (with disconnected superconducting 
regions) and high-resistive core. The measured conductivity of these 
composite system should be much higher than ``intrinsic'' conductivity 
of the core. The increasing $U$ leads to the increase in the 
current and this must induce the depression of surface superconductivity 
and, hence, the increasing of the sample resistance. This corresponds to the 
observed $R(U)$ behavior in low-voltage range (Figs. 3, 5 and~6). 
\par 
One of the obvious reasons for the superconductivity depression at increasing
$U$ is the increasing in current density (this leads to reducing in $T_{c}$). 
It must not be ruled out, however, in this case the possible influence of 
Joule heating in low-resistive surface layer on the conductivity of whole
system since the Joule heat (as well as current) is much more in this layer
than in the core. It is known that Joule heating plays a crucial role in
the breakdown of superconductivity in composite or inhomogeneous 
superconductors \cite{gur}. The Joule heating may result (among other things)
in resistive domains and negative differential conductance \cite{gur}. 
The latter can be actually seen in the measured $I$--$V$ characteristics 
at low enough temperatures (Fig. 5). It can not be excluded that 
the observed negative differential conductance is connected with some 
of the described in Ref. \cite{gur} mechanisms of heat breakdown of 
superconductivity. The results obtained do not provide reason enough to 
consider this question in detail. In any case, however, we believe that 
increasing $U$ leads to superconductivity depression and, hence, to the 
resistance increase. 
\par
The magnetic field should also reduce the superconductivity. In this 
connection the observed positive MR  at low-voltage range and the transition 
to negative MR with increasing~$U$ (Fig. 9) can be considered as the 
important argument to support the existence of oxygen-enriched surface layer 
(with superconducting inclusions) in the sample. 
A close look at Fig.~9c shows that when temperature drops, the MR is first 
negative and then becomes postive. It is significant that the positive MR is 
combined with {\it decreasing} of the resistance with decreasing temperature 
at $H = 0$ whereas the negative MR is combined with {\it increasing} 
of resistance as the temperature decreases. The resistance decrease with 
decreasing temperature takes place only at low-voltage range 
where surface superconductivity is not depressed (Fig.~9, see also Fig.~4). 
This decrease can be explained by enhancing of Josephson 
coupling within some confined groups of superconducting 
regions with  decreasing temperature. Such effect is quite typical for 
granular metals in which the competition of the hopping conduction and 
Josephson coupling takes place \cite{bel4}. All these effects (especially, 
the positive MR 
combined with resistance decreasing with decreasing temperature) can be 
considered as a direct evidence of superconductivity effect in sample studied.
\par
It is reasonable to expect that at high enough voltage the surface 
superconductivity will be depressed completely after which the 
non-linear conductance of the whole system would change over to Ohmic 
behavior (Fig. 5 and 6).  Only at highest applied voltage the non-linear 
behavior appears again for the reasons which we have mentioned above.
\par
Above $T \simeq 20$~K, where the superconductivity effect should not 
take place, the non-linear behavior of conduction at low-voltage range 
still remains. 
It is weaker than at $T~<~20$~K, and appears in radically changed form: 
the resistance decreases with increasing $U$ (Figs. 6 and~7), at high 
enough voltage the resistance seems to saturate, that is, the transition to 
Ohmic behavior occurs (Fig. 5). This type of 
non-linearity can also be adequately explained in the context of our main 
conjecture (oxygen-enriched surface layer). The low-resistive surface layer 
is inhomogeneous. It consists of disconnected (dispersed) high-conducting 
regions in dielectric matrix. Generally the surface layer would constitute a 
percolation system with tunneling (or hopping) between disconnected 
conducting regions. It is just the tunneling which is responsible for the 
non-linearity of this type of composite system \cite{sen,gupta}. The 
distinctive feature of these systems is the transition from non-Ohmic to 
Ohmic behavior of conduction at increasing applied electric field (or 
temperature). The transition of this type was observed on Ag particles in 
KCl matrix \cite{chen} and in a semicontionuous gold film near the 
percolation threshold \cite{bel3}. In Ref. \cite{bel3} such behavior was
attributed (in line with theory of Ref. \cite{mos}) to increasing of 
the probability of tunneling with increasing applied voltage $U$ or 
temperature. The percolation approach of Refs. \cite{sen,gupta} leads to 
essentially same result. Thus we believe that the observed change-over from 
non-linear conductance to Ohmic behavior at low-voltage range (Fig. 5-7)
with increasing $U$ is connected with percolating structure of 
oxygen-enriched surface layer and should be attributed to theoretical 
mechanisms similar those of Refs. \cite{sen,gupta,mos}. Once the conduction
of this layer becomes Ohmic beyond some voltage, the behavior of the whole 
sample becomes also Ohmic up to the highest voltage, where the influence of 
electric field (or Joule heating) on hopping conduction becomes perceptible. 
\par
In conclusion, it may be said that our conjection about the oxygen-enriched 
surface layer enables us to explain all the observed unusual non-linear 
effects and magnetoresistance behavior of studied sample of 
La$_{2}$CuO$_{4 + \delta}$. The results obtained demonstrate that transport
properties of cuprate oxides may be determined in essential degree by 
structural or stoichimetric inhomogeneities. This circumstance should be 
taken into account at evaluation of ``quality'' of high-temperature 
superconductors on the basis of transport properties.            
\acknowledgments              
We are very grateful to S. I. Shevchenko for critical reading of the 
manuscript and helpful comments.

\begin{figure}
\caption{Temperature dependences of magnetic susceptibility $\chi(T)$ in 
the magnetic field $H = 0.83$~T of the single-crystal sample of the 
La$_{2}$CuO$_{4 + \delta}$ in the initial state ($\bullet$) and 
after the outlined heat treatment in helium and air ($\circ$). 
The quantity $\chi_{c}$ 
corresponds to the measurements in a magnetic field parallel to the 
crystallographic axis $c$ (the unit cell corresponds 
to the crystallographic $Bmab$ structure 
in which $a < b < c$, $c$ being the tetragonal axis). 
The dependences $\chi_{c}(T)$ are presented by two upper curves. The
quantity $\chi_{ab}$ is the susceptibility in a magnetic field parallel to 
Cu-O planes (two bottom curves). The positions of maximums of $\chi(T)$
dependences correspond to N\'eel temperature $T_{N}$.    
}
\label{Fig.1}
\end{figure}

\begin{figure}
\caption{The dependences of resistance $R$ (on logarithmic scale) on 
$T^{-1/4}$ of the sample in initial state (curve 1) and after the outlined 
heat treatment in helium and air (curve 2). The dependences were registered 
at applied voltage $U = 25$~V.
}
\label{Fig.2}
\end{figure}

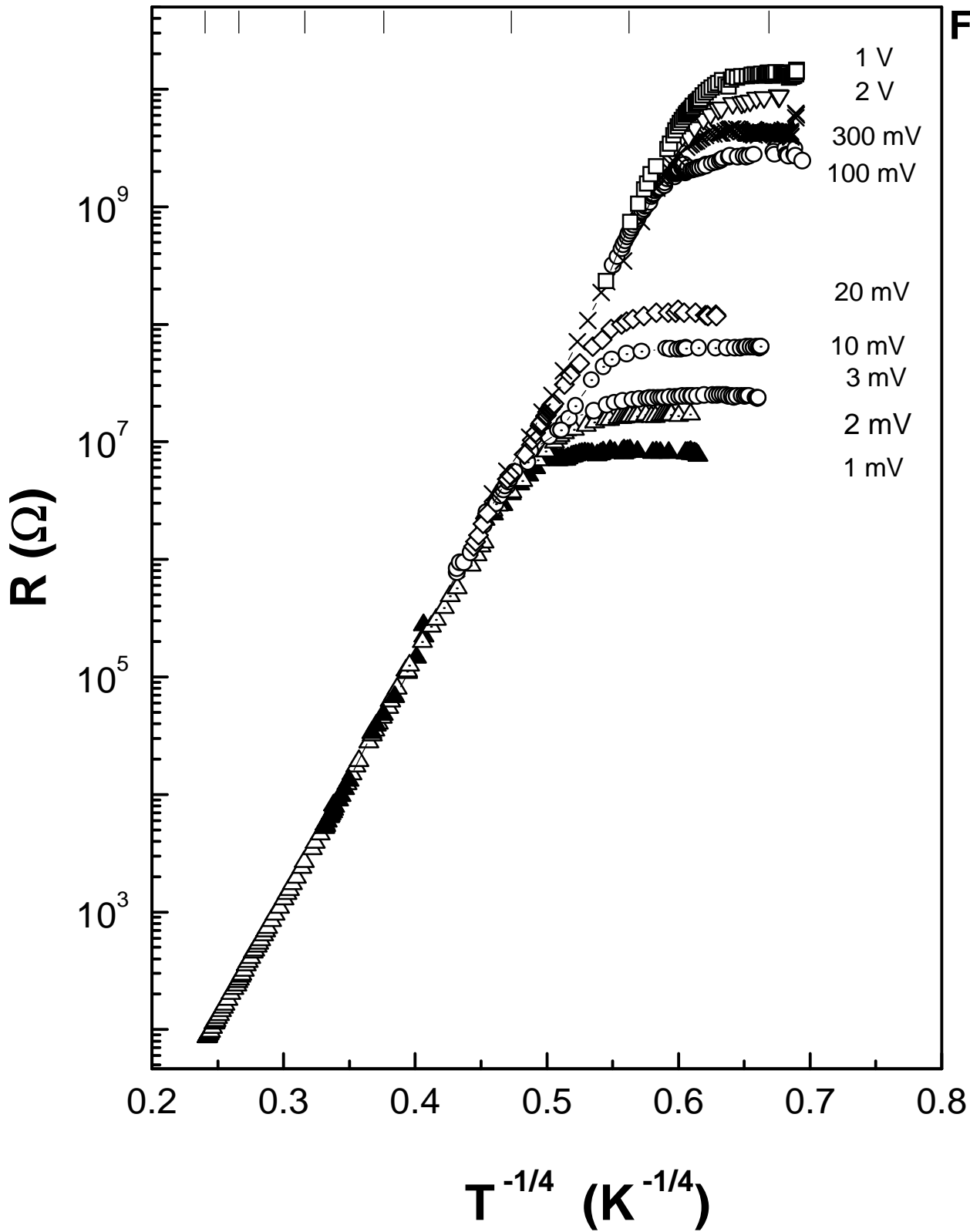
\begin{figure}
\caption{The dependences of resistance $R$ (on logarithmic scale) on 
$T^{-1/4}$ of the sample studied at different magnitudes of applied
voltage.
}
\label{Fig.3}
\end{figure}

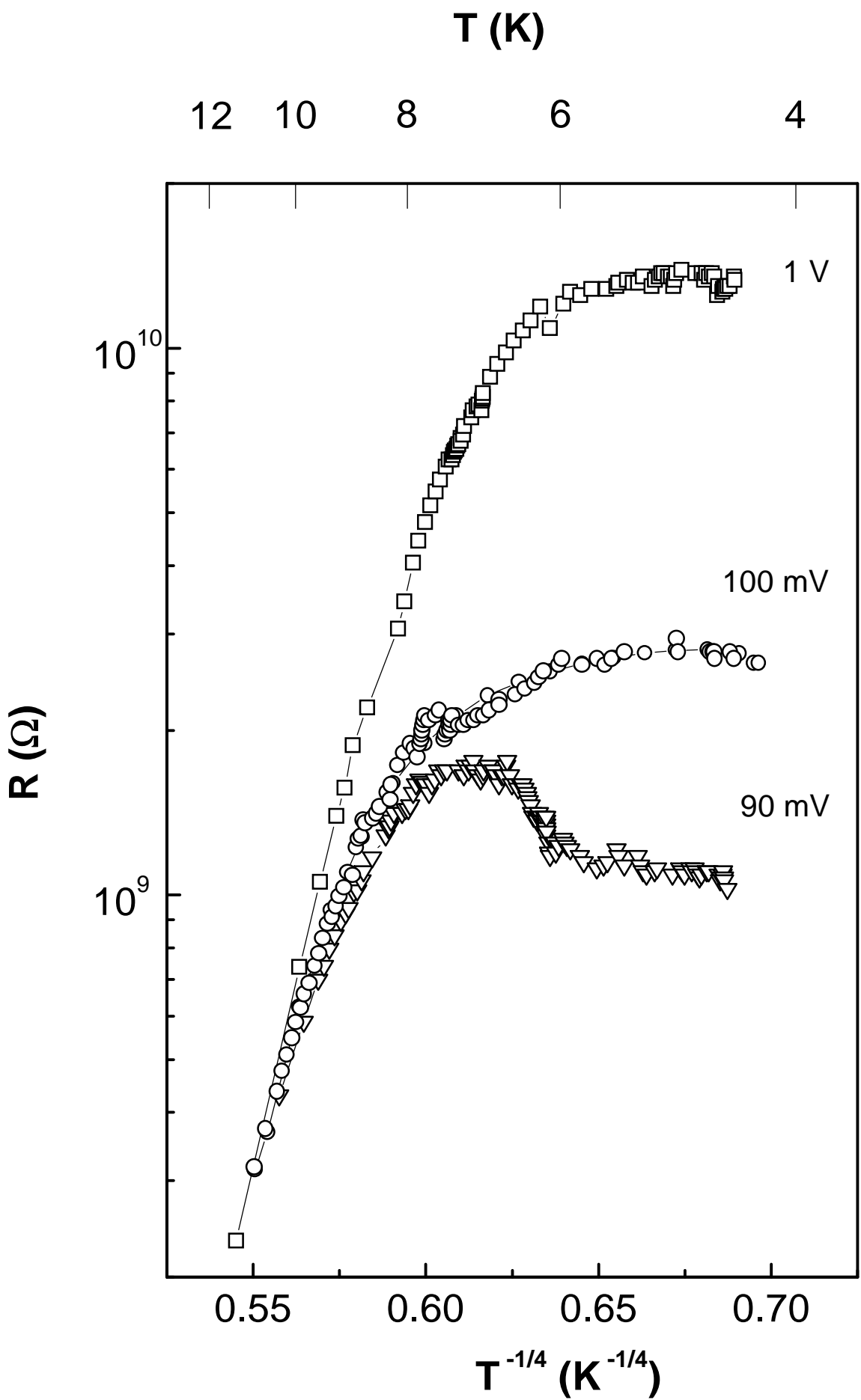
\begin{figure}
\caption{A selection of dependences of resistance $R$ (on logarithmic scale) 
on $T^{-1/4}$ presented on an enlarged scale as compared with 
Fig.~3. They demostrate the peculiarities of $R(T)$ behavior at low 
temperature range at different magnitudes of applied voltage. It can be seen 
that at high voltage the resistance saturates with decreasing temperature, 
but at low enough voltage it {\it decreases} with decreasing temperature.   
}
\label{Fig.4}
\end{figure}

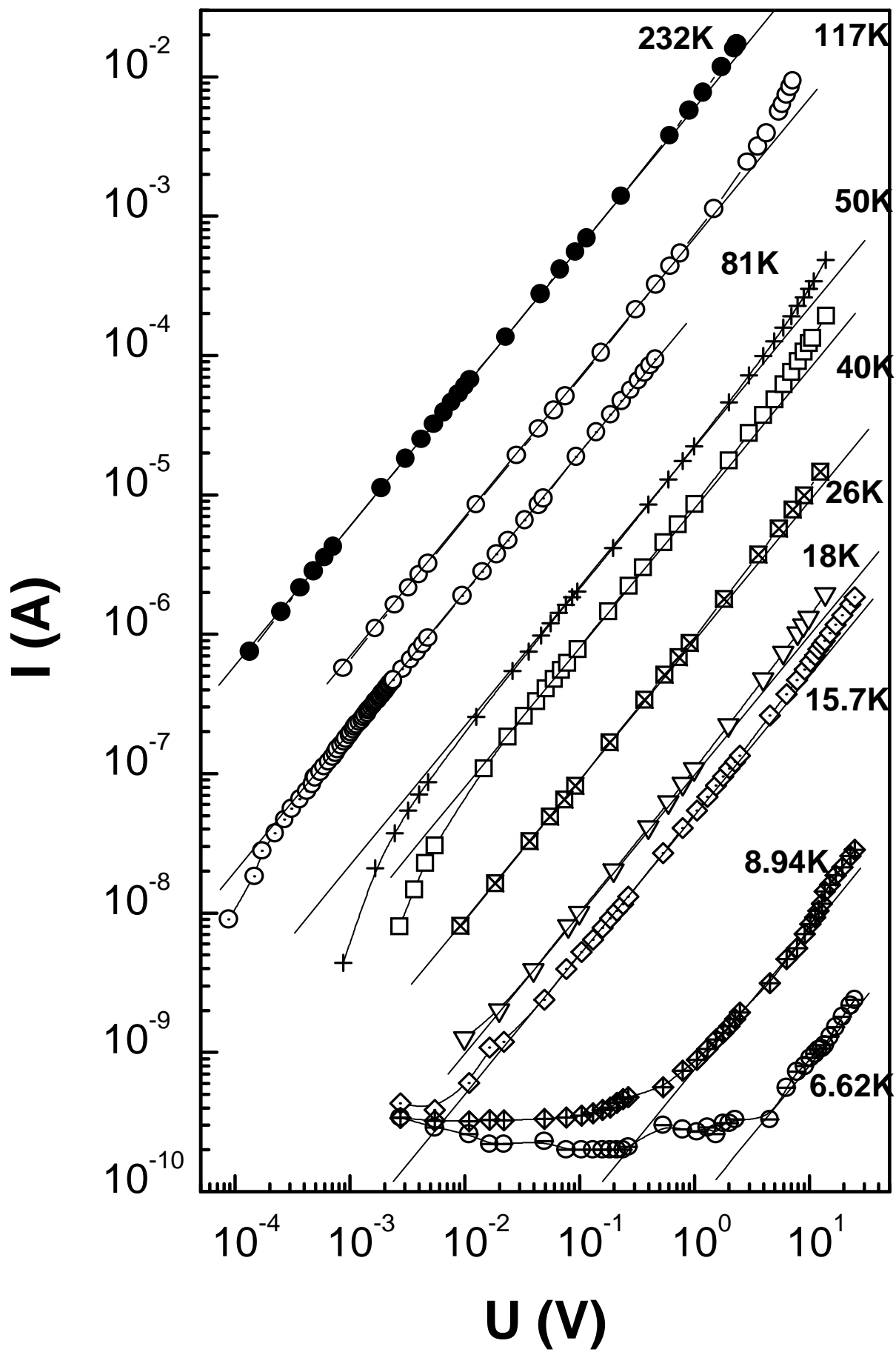
\begin{figure}
\caption{A set of $I-V$ curves (in logarithmic coordinates) for different 
temperatures.
}
\label{Fig.5}
\end{figure}

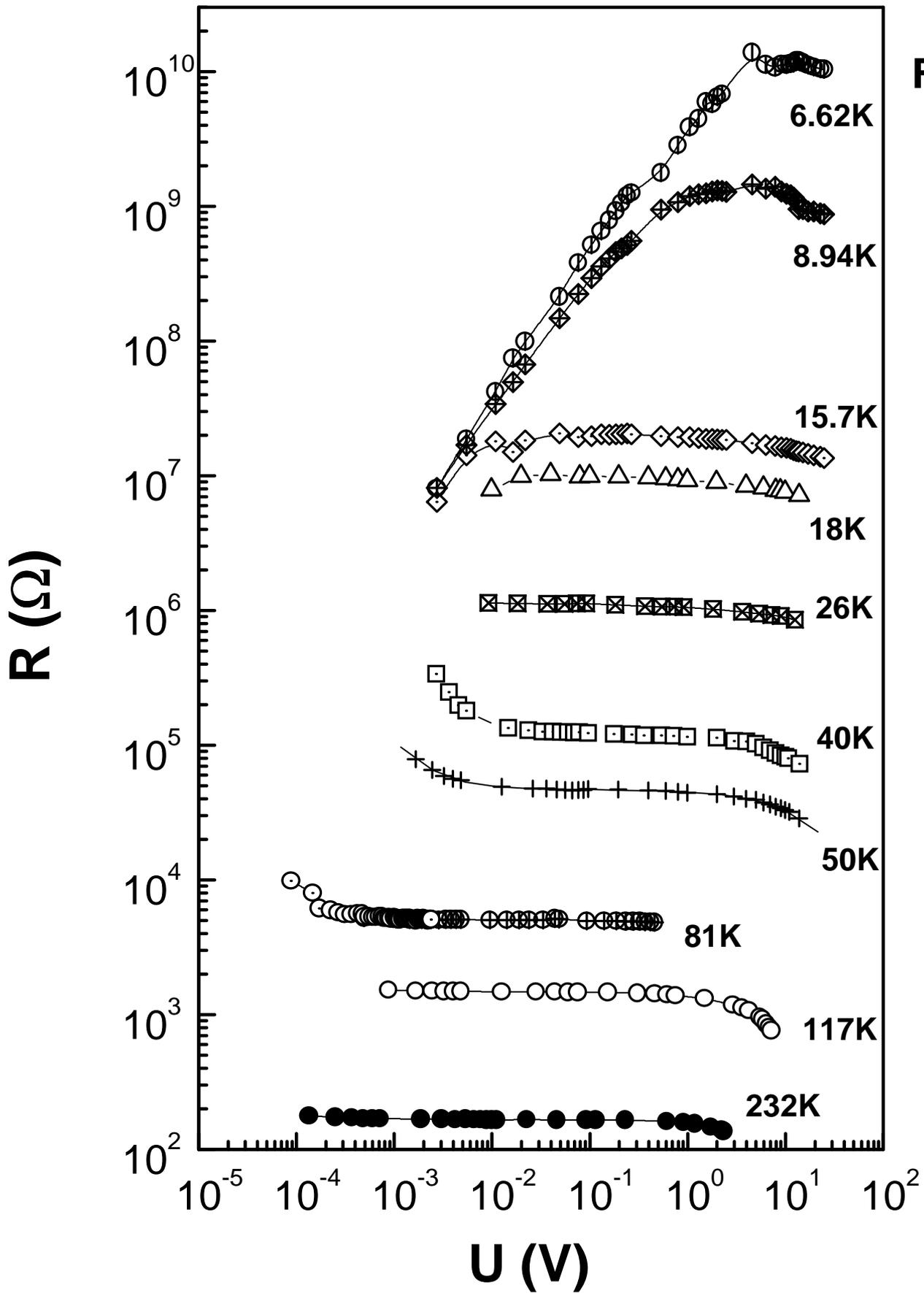
\begin{figure}
\caption{A set of voltage dependences of resistance $R$ 
(in logarithmic coordinates) for the same temperatures as in Fig.~3.
}
\label{Fig.6}
\end{figure}

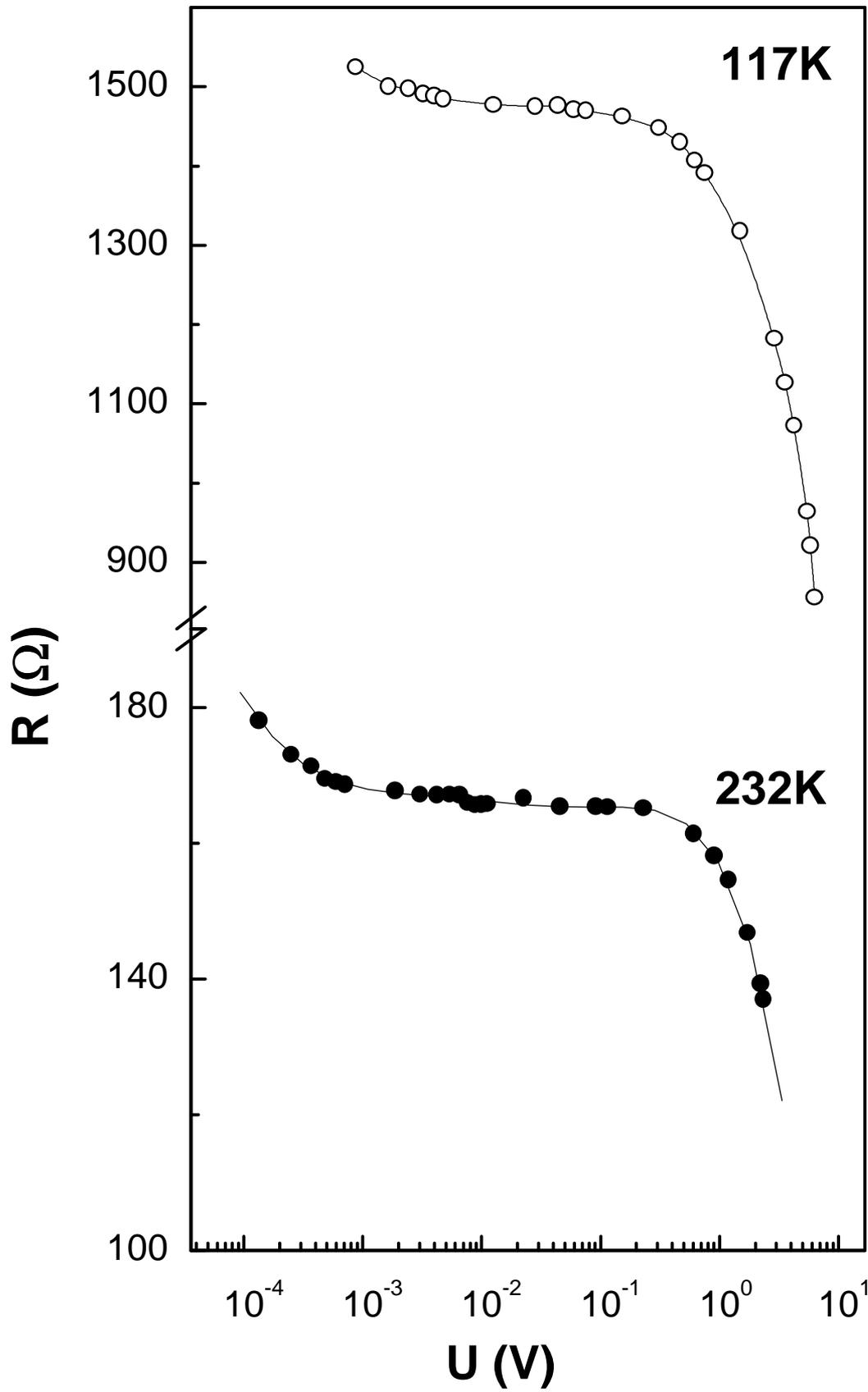
\begin{figure}
\caption{The semilogarithmic plots of voltage dependences of 
resistance $R$ for two temperatures above 100 K.
}
\label{Fig.7}
\end{figure}

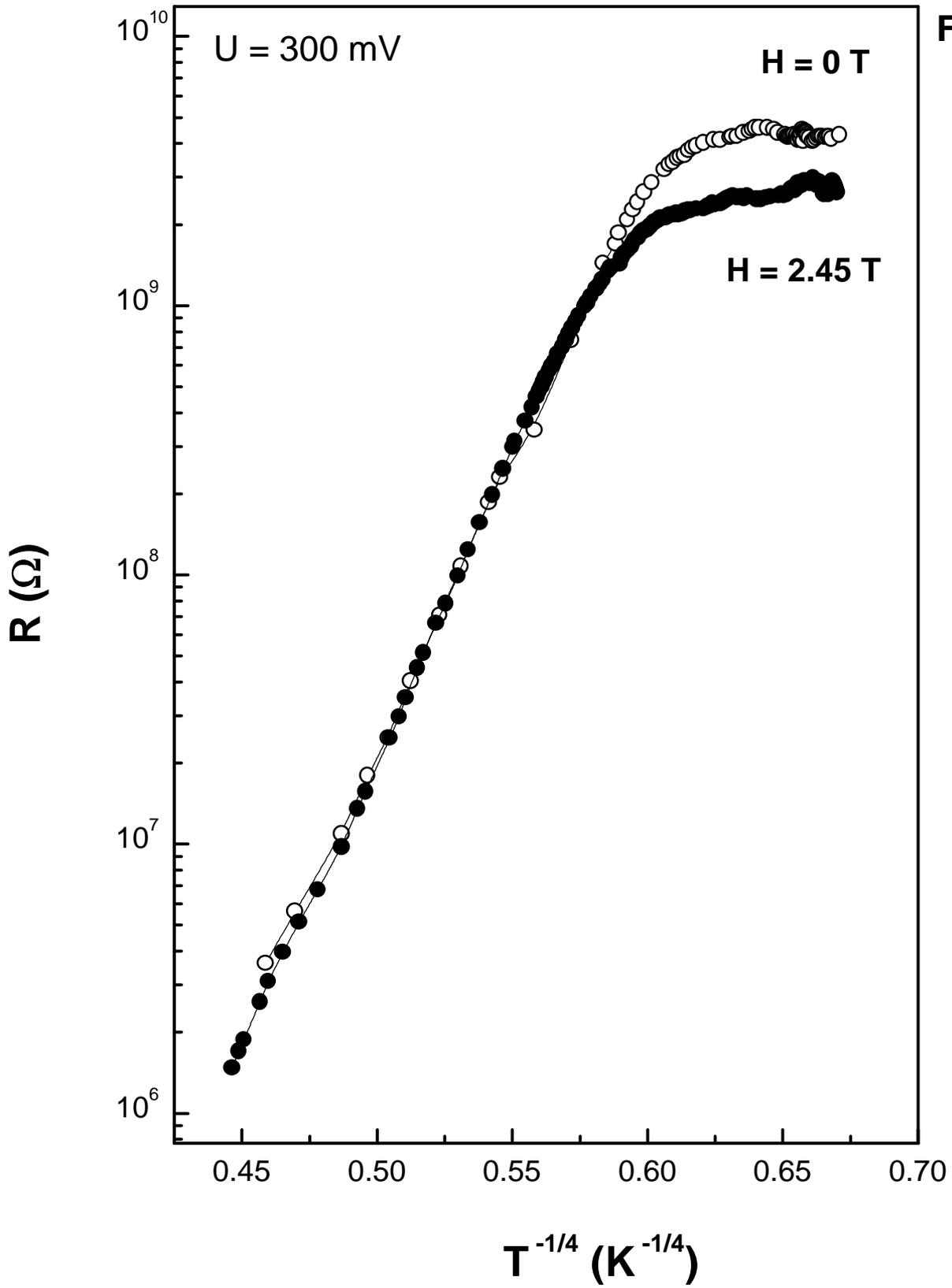
\begin{figure}
\caption{The dependences of resistance $R$ (on logarithmic scale) on 
$T^{-1/4}$ at $U = 300$~mV registered in a magnetic field $H = 0$ and
$H = 2.45$~T.
}
\label{Fig.8}
\end{figure}

\begin{figure}
\caption{The temperature dependences of resistance $R(T)$ at $H = 0$~T and
at some constant magnitudes of $H$. These dependences were registered at 
different applied voltage $U$: 1 V (a); 100~mV~(b); 70 mV (c).
}
\label{Fig.9}
\end{figure}


\begin{references}
\bibitem[*]{byline}E-mail: belevtsev@ilt.kharkov.ua
\bibitem{ps}E. Sigmund and K. A. M\"{u}ller (Eds.), {\it Phase separation
in Cuprate Superconductors,} Springer-Verlag, Heidelberg (1994).
\bibitem{browning}V. M. Browning, E. F. Skelton, M. S. Osofsky,
S. B. Qadri, J. Z. Hu, L. W. Finger, and P. Caubet, Phys. Rev. 
{\bf B56,} 2860 (1997).
\bibitem{john}D. C. Johnston, J. P. Stokes, D. P. Goshorn, and 
J. T. Lewandowski, Phys. Rev. {\bf B36,} 4007 (1987).
\bibitem{birg}R. J. Birgeneau and G. Shirane, in 
{\it Physical Properties of High Temperature Superconductors I},
[ed. by D. M. Ginsberg], World Scientific, Singapore (1989), 
Ch. 4, pp.151-211.
\bibitem{kremer}R. K. Kremer, A. Simon, E. Sigmund, and V. Hizhnyakov,
in {\it Phase Separation in Cuprate Superconductors}
[ed. by E. Sigmund and K. A. M\"{u}ller], Springer-Verlag, Heidelberg 
(1994), pp. 66-81.
\bibitem{sato} H. Sato, M. Naito, and H. Yamamoto, Physica 
{\bf C280,} 178 (1997).
\bibitem{jorg}J. D. Jorgensen, B. Dabrowski, S. Pei, D. G. Hinks,
L. Soderholm, B. Morosin, J. E. Schirber, E. L. Venturini, and 
D. S. Ginley, Phys. Rev. {\bf B38,} 11337 (1988).  
\bibitem{hund}M. F. Hundley, R. S. Kwok, S.-W. Cheong, J. D. Thompson, 
and Z. Fisk, Physica {\bf C172,} 455 (1991). 
\bibitem{ryder}J. Ryder, P. A. Midgley, R. J. Beynon, D. L. Yates,
L. Afalfiz, and J. A. Wilson, Physica {\bf C173,} 9 (1991)
\bibitem{zakh}A. A. Zakharov and A. A. Nikonov, JETP Lett. {\bf 60,}
348 (1994).
\bibitem{katya}E. L. Vavilova, N. N. Garif'yanov, E. F. Kukovitsky, 
G. B. Teitel'baum, Physica {\bf C264,} 74 (1996).
\bibitem{kast}M. A. Kastner, R. J. Birgeneau, C. Y. Chen, Y. M. Chiang,
D. R. Gabbe, H. P. Jenssen, T. Junk, C. J. Peters, P. J. Picone, Tineke Thio,
T. R. Thurston, and H. L. Tuller, Phys. Rev. {\bf B37,} 111 (1988).
\bibitem{zakh2}A. A. Zakharov, E. P. Krasnoperov, B. I. Savel'ev,
A. A. Teplov, M. B. Tsetlin, A.~A.~Shikov, Sverkhprovodimost': Fiz., Khim.,
Tekh. {\bf 4,} 1906 (1991). 
\bibitem{bel1}B. I. Belevtsev, N. V. Dalakova, and A. S. Panfilov,
Low Temp. Phys. {\bf 23,} 274 (1997).
\bibitem{bel2}B. I. Belevtsev, N. V. Dalakova, and A. S. Panfilov,
Physica {\bf C282-287,} 1223 (1997). 
\bibitem{mott}N. F. Mott and E. A. Davis, {\it Electron Processes in
Noncrystalline Materials,} Clarendon Press, Oxford (1979).
\bibitem{iye}Y. Iye,  in 
{\it Physical Properties of High Temperature Superconductors III},
[ed. by D. M. Ginsberg], World Scientific, Singapore (1992), 
Ch. 4, pp.285-361.
\bibitem{forro}L. Forro, Int. J. of Mod. Phys. {\bf 8,} 829 (1994).
\bibitem{loktev}V. M. Loktev, Fiz. Nizk. Temp. {\bf 22,} 3 (1996) 
[Low Temp. Phys. {\bf 22,} 1 (1996)].
\bibitem{chen}C. Y. Chen, R. J. Birgeneau, M. A. Kastner, N. W. Preyer,
and T. Thio, Phys. Rev. {\bf B43,} 392 (1991).
\bibitem{thio}T. Thio, C. Y. Chen, B. S. Freer, D. R. Gable, 
H. P. Jenssen, M. A. Kastner, P. J. Picone, and N. W. Preyer, 
Phys. Rev. {\bf B41,} 231 (1990)
\bibitem{shklo}B. I. Shklovskii and  A. L. Efros, 
{\it Electronic  Properties of Doped Semiconductors,} 
Springer-Verlag, New York (1984).
\bibitem{birg2}R. J. Birgeneau, F. C. Chou, Y. Endoh, M. A. Kastner,
Y. S. Lee, G. Shirane, J. M. Tranquad, B. O. Wells, and K. Yamada,
in {\it Proc. of the 10th Anniversary HTS Workshop on Physics, Materials
and Applications, March 12--16, 1996, Houston, Texas, USA},
[ed. by B. Batlog, C. A. Chu, W. K. Chu, D. U. Gubster, and 
K. A. M\"{u}ller], World Scientific, Singapore (1996), pp. 421-424.
\bibitem{gur}A. V. Gurevich, R. G. Mints, and A. L. Rakhmanov, {\it
Fizika kompozitnykh svekhprovodnikov (Physics of composite 
superconductors),} Nauka, Moscow (1987).
\bibitem{bel4}B. I. Belevtsev, Sov. Phys. Uspekhi {\bf 33,} 36 (1990).
\bibitem{sen}A. K. Sen and A. Kar Gupta, in {\it Non-linearity and 
Breakdown in Soft Condensed Matter}, [ed. by K. K. Bardhan, B. K. 
Chakrabarti and A. Hansen], {\it Lecture Notes in Physics}, {\bf 437},
Springer-Verlag, Berlin (1994), pp.271-287.
\bibitem{gupta} A. Kar Gupta and A. K. Sen, Phys. Rev. {\bf B57,}
3375 (1998).
\bibitem{chen}I-G. Chen and W. B. Johnson, J. Mat. Sci. {\bf 27,} 5497 (1992).
\bibitem{bel3}B. I. Belevtsev, E. Yu. Belyaev, Yu. F. Komnik, and
E. Yu. Kopeichenko, Low Temp. Phys. {\bf 23,} 724 (1997).
\bibitem{mos}M. Mostefa, D. Bourbie, and G. Olivier, Physica
{\bf B160,} 186 (1989).
\end{references}
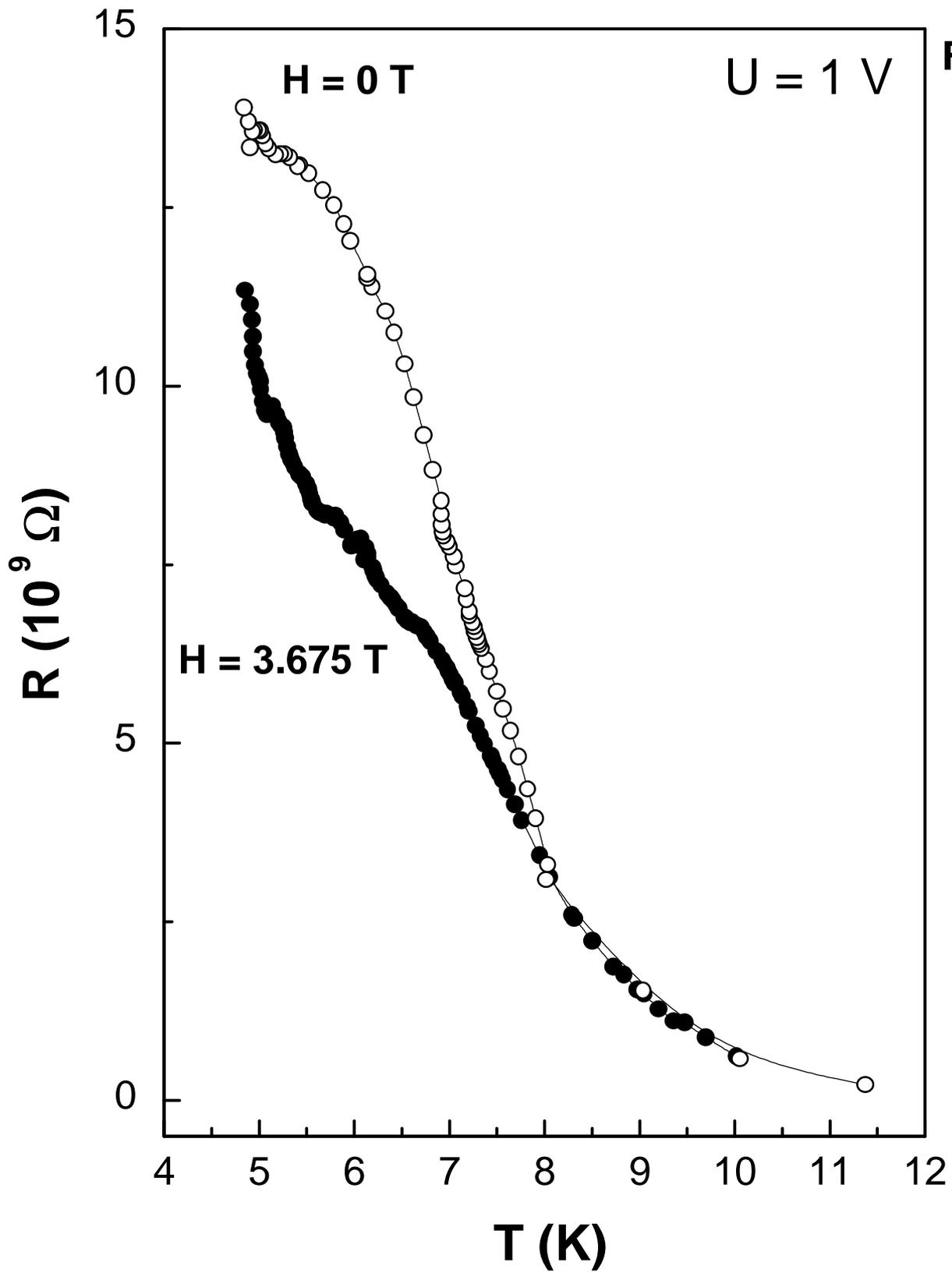
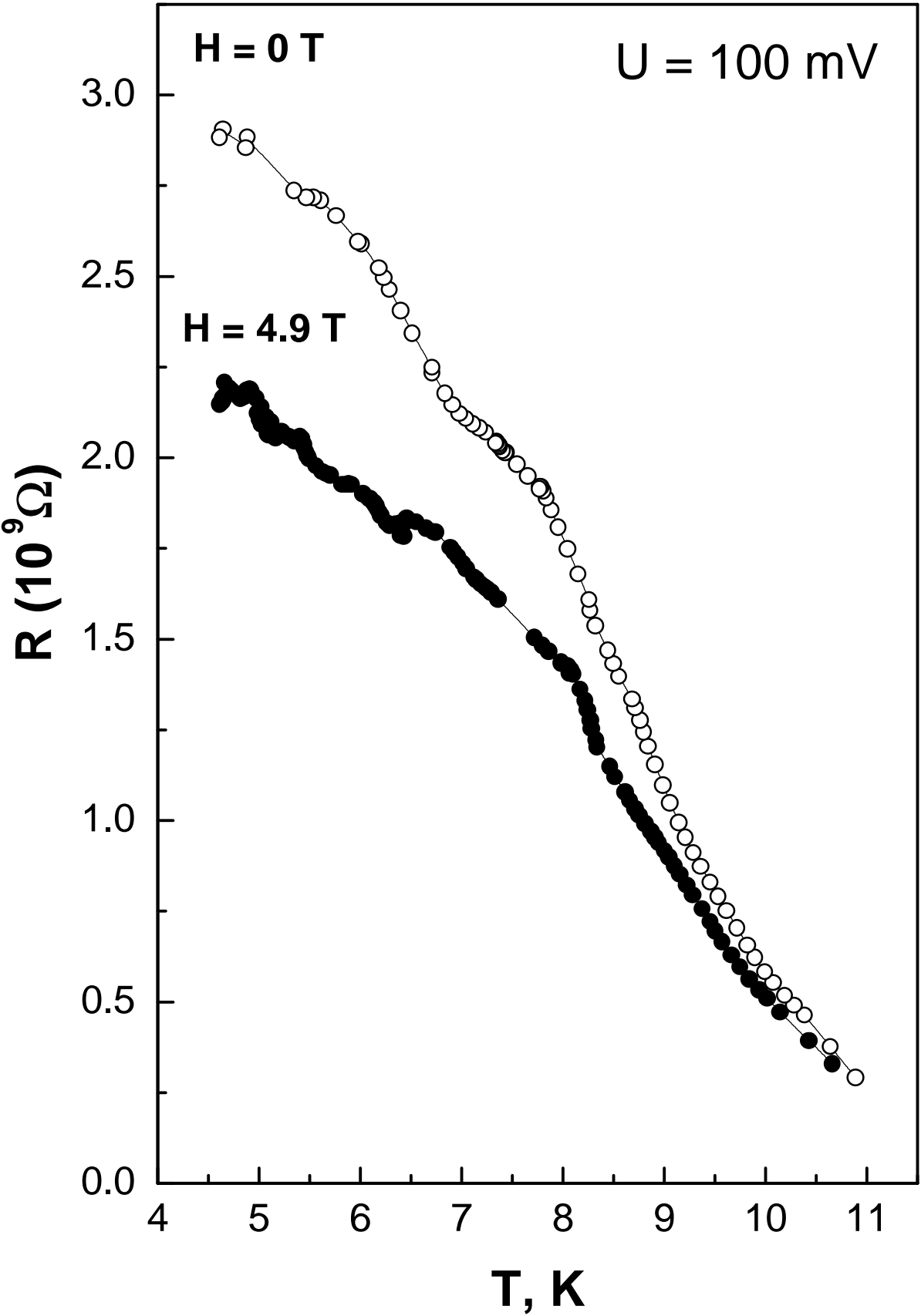
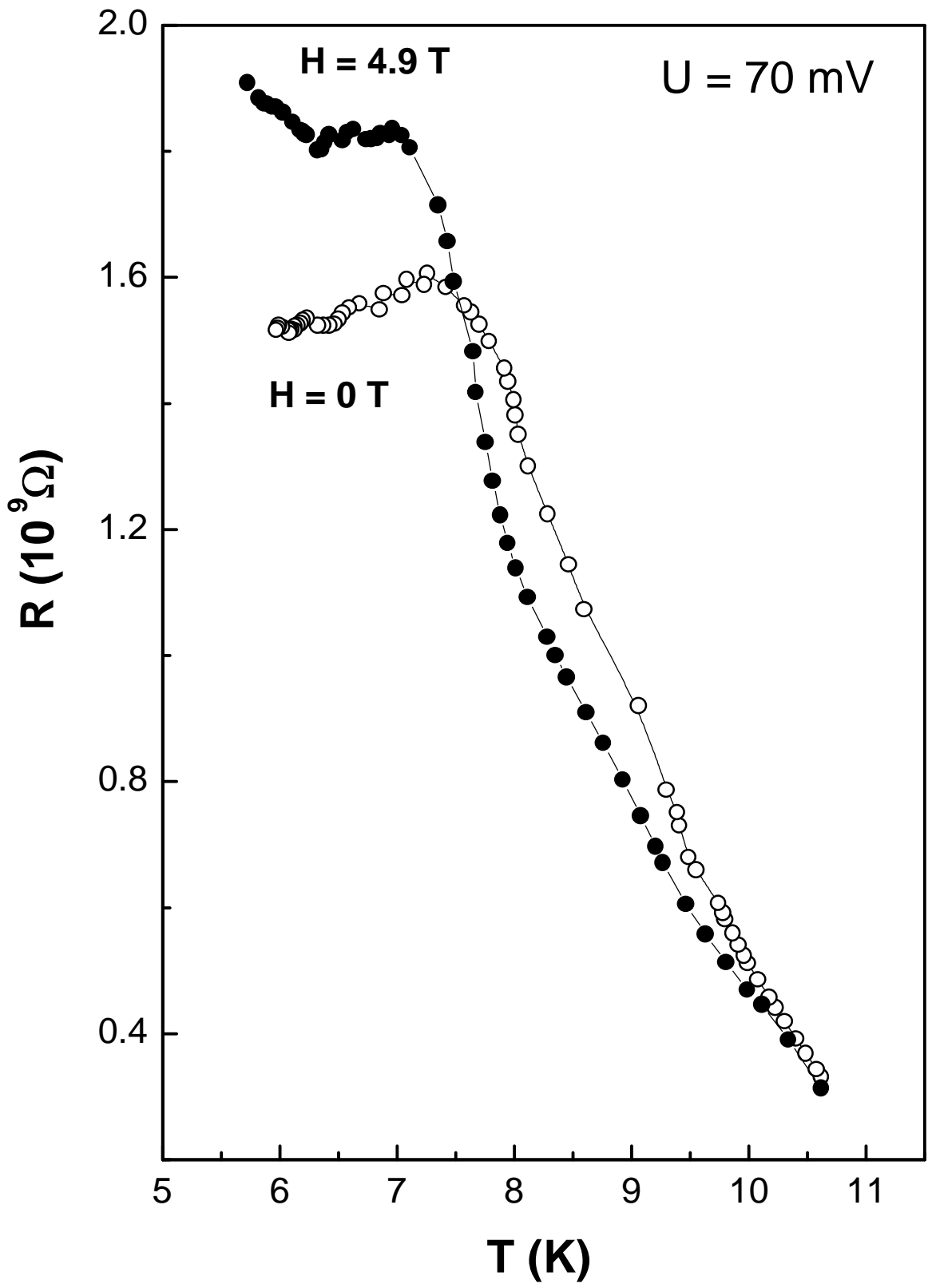
\end{document}